\def\DeclareAbbreviation#1#2{%
   \DeclareRobustCommand*#1{\@journalname{#2}}}
\def\@journalname#1{{\normalfont#1}}
\DeclareAbbreviation\aj{AJ}
\DeclareAbbreviation\araa{ARA\&A}
\DeclareAbbreviation\apj{ApJ}
\DeclareAbbreviation\apjl{ApJL}
\DeclareAbbreviation\apjs{ApJS}
\DeclareAbbreviation\ao{Appl.\ Opt.}
\DeclareAbbreviation\apss{Ap\&SS}
\DeclareAbbreviation\aap{A\&A}
\DeclareAbbreviation\aapr{A\&AR}
\DeclareAbbreviation\aaps{A\&AS}
\DeclareAbbreviation\azh{AZh}
\DeclareAbbreviation\baas{BAAS}
\DeclareAbbreviation\jrasc{JRASC}
\DeclareAbbreviation\memras{MmRAS}
\DeclareAbbreviation\mnras{MNRAS}
\DeclareAbbreviation\pra{Phys.\ Rev.\ A}
\DeclareAbbreviation\prb{Phys.\ Rev.\ B}
\DeclareAbbreviation\prc{Phys.\ Rev.\ C}
\DeclareAbbreviation\prd{Phys.\ Rev.\ D}
\DeclareAbbreviation\pre{Phys.\ Rev.\ E}
\DeclareAbbreviation\prl{Phys.\ Rev.\ Lett.}
\DeclareAbbreviation\pasp{PASP}
\DeclareAbbreviation\pasj{PASJ}
\DeclareAbbreviation\qjras{QJRAS}
\DeclareAbbreviation\skytel{S\&T}
\DeclareAbbreviation\solphys{Sol.\ Phys.}
\DeclareAbbreviation\sovast{Soviet\ Ast.}
\DeclareAbbreviation\ssr{Space\ Sci.\ Rev.}
\DeclareAbbreviation\zap{ZAp}
\DeclareAbbreviation\nat{Nature}
\DeclareAbbreviation\iaucirc{IAU\ Circ.}
\DeclareAbbreviation\aplett{Astrophys.\ Lett.}
\DeclareAbbreviation\apspr{Astrophys.\ Space\ Phys.\ Res.}
\DeclareAbbreviation\bain{Bull.\ Astron.\ Inst.\ Netherlands}
\DeclareAbbreviation\fcp{Fund.\ Cosmic\ Phys.}
\DeclareAbbreviation\gca{Geochim.\ Cosmochim.\ Acta}
\DeclareAbbreviation\grl{Geophys.\ Res.\ Lett.}
\DeclareAbbreviation\jcp{J.\ Chem.\ Phys.}
\DeclareAbbreviation\jgr{J.\ Geophys.\ Res.}
\DeclareAbbreviation\jqsrt{J.\ Quant.\ Spectrosc.\ Radiat.\ Transfer}
\DeclareAbbreviation\memsai{Mem.\ Soc.\ Astron.\ Italiana}
\DeclareAbbreviation\nphysa{Nucl.\ Phys.\ A}
\DeclareAbbreviation\physrep{Phys.\ Rep.}
\DeclareAbbreviation\physscr{Phys.\ Scr.}
\DeclareAbbreviation\planss{Planet.\ Space\ Sci.}
\DeclareAbbreviation\procspie{Proc.\ SPIE}
\DeclareAbbreviation\aip{AIP Conf.\ Proc.}
\DeclareAbbreviation\asp{ASP Conf.\ Ser.}

\documentclass[a4paper,12pt]{spieman}  
\usepackage{amsmath}  
\usepackage{graphicx}
\usepackage{amssymb}
\usepackage{setspace}
\usepackage{tocloft}
\usepackage{color}

\title{MuSCAT2: four-color simultaneous camera for the 1.52-m Telescopio Carlos S\'anchez} 

\author{Norio Narita,\supscr{a,b,c,d,e}
Akihiko Fukui,\supscr{e,f,g}
Nobuhiko Kusakabe,\supscr{b,d}
Noriharu Watanabe,\supscr{d,h}\\
Enric Palle,\supscr{e,i}
Hannu Parviainen,\supscr{e,i}
Pilar Monta\~n\'es-Rodr\'iguez,\supscr{e,i}
Felipe Murgas,\supscr{e,i}\\
Matteo Monelli,\supscr{e,i}
Marta Aguiar,\supscr{e}
Jorge Andres Perez Prieto,\supscr{e,i}
\'Alex Oscoz,\supscr{e,i}\\
Jerome de Leon,\supscr{a}
Mayuko Mori,\supscr{a}
Motohide Tamura,\supscr{a,b,d}
Tomoyasu Yamamuro,\supscr{j}\\
Victor J. S. B\'{e}jar,\supscr{e,i}
Nicolas Crouzet,\supscr{e,i}
Diego Hidalgo,\supscr{e,i}
Peter Klagyivik,\supscr{e,i}\\
Rafael Luque,\supscr{e,i}
Taku Nishiumi\supscr{k}
}

\affiliation{\supscrsm{a}Department of Astronomy, The University of Tokyo, 7-3-1 Hongo, Bunkyo-ku, Tokyo 113-0033, Japan\\
\supscrsm{b}Astrobiology Center, 2-21-1 Osawa, Mitaka, Tokyo 181-8588, Japan\\
\supscrsm{c}JST, PRESTO, 7-3-1 Hongo, Bunkyo-ku, Tokyo 113-0033, Japan\\
\supscrsm{d}National Astronomical Observatory of Japan, 2-21-1 Osawa, Mitaka, Tokyo 181-8588, Japan\\
\supscrsm{e}Instituto de Astrof\'{i}sica de Canarias (IAC), 38205 La Laguna, Tenerife, Spain\\
\supscrsm{f}Department of Earth and Planetary Science, The University of Tokyo, 7-3-1 Hongo, Bunkyo-ku, Tokyo 113-0033, Japan\\
\supscrsm{g}Subaru Telescope Okayama Branch Office, National Astronomical Observatory of Japan, 3037-5 Honjo, Kamogata, Asakuchi, Okayama 719-0232, Japan\\
\supscrsm{h}SOKENDAI (The Graduate University of Advanced Studies), 2-21-1 Osawa, Mitaka, Tokyo 181-8588, Japan\\
\supscrsm{i}Departamento de Astrof\'{i}sica, Universidad de La Laguna (ULL), 38206 La Laguna, Tenerife, Spain\\
\supscrsm{j}OptCraft, 3-16-8 Higashi-Hashimoto, Midori-ku, Sagamihara, Kanagawa 252-0144, Japan\\
\supscrsm{k}Department of Physics, Kyoto Sangyo University, Motoyama, Kamigamo, Kita-ku, Kyoto, 603-8555 Japan}

\cftpagenumbersoff{figure}
\cftpagenumbersoff{table} 
\begin{document} 
\maketitle 

\begin{abstract}
We report the development of a 4-color simultaneous camera for the 1.52~m 
Telescopio Carlos S\'anchez (TCS) in the Teide Observatory, Canaries, Spain.
The new instrument, named MuSCAT2, has a capability of 4-color simultaneous imaging
in $g$ (400--550 nm), $r$ (550--700 nm), $i$ (700--820 nm),  and $z_s$ (820--920 nm) bands.
MuSCAT2 equips four 1024$\times$1024 pixel CCDs, having
a field of view of 7.4$\times$7.4 arcmin$^2$ with a pixel scale of 0.44 arcsec per pixel.
The principal purpose of MuSCAT2 is to perform high-precision multi-color exoplanet transit photometry.
We have demonstrated photometric precisions of
0.057\%, 0.050\%, 0.060\%, and 0.076\% as root-mean-square residuals of 60~s binning 
in $g$, $r$, $i$ and  $z_s$ bands, respectively, for a G0 V star WASP-12 ($V=11.57\pm0.16$).
MuSCAT2 has started science operations since January 2018,
with over 250 telescope nights per year.
MuSCAT2 is expected to become a reference tool for exoplanet transit observations,
and will substantially contribute to the follow-up of the TESS and PLATO space missions.  
\end{abstract}

\keywords{instrumentation, exoplanets, multicolor, photometry, transits.}

{\noindent \footnotesize{\bf Address all correspondence to}: Norio Narita, The University of Tokyo, \\ 7-3-1 Hongo, Bunkyo-ku, Tokyo 113-0033, Japan; Tel: +81 3-5841-1032; \\
E-mail:  narita@astron.s.u-tokyo.ac.jp }

\begin{spacing}{1}   

\section{Introduction}
\label{sect:intro}  

Transiting planets, periodically passing in front of their host stars, are
valuable targets for exoplanet studies, since one can investigate
the true mass, radius, density, orbital obliquity, and atmosphere of such planets.
The number of discovered transiting planets are drastically increasing in recent years
thanks to intense transit surveys from the ground
\cite{2007ApJ...656..552B,2007MNRAS.375..951C,2012ApJ...761..123S,2009Natur.462..891C,2016Natur.533..221G,2018MNRAS.475.4476W}
and space\cite{2006cosp...36.3749B,2010Sci...327..977B,2014PASP..126..398H}\,\!.
Moreover, the NASA's new mission {\it TESS}\cite{2015JATIS...1a4003R} has started operations in 2018, and 
the future ESA mission {\it PLATO}\cite{2014ExA....38..249R} is planned to be launched around 2026.
Those missions will especially focus on relatively bright nearby planet host stars.
Hence it is expected that hundreds or thousands of new transiting planets, suitable for
further characterizations, will be discovered in the vicinity of our Solar system in the near future.

On the other hand, discovered candidates of transiting planets are not always {\it bona fide} planets.
This is because eclipsing binaries may mimic transit-like signals in photometric survey data.
False positive rates are especially worse for ground-based surveys (e.g., more than 98\% for
the recent KELT survey\cite{2018arXiv180301869C}),
and still not negligible even for a space-based survey
like {\it Kepler}\cite{2012A&A...545A..76S}\,\!.
This is also true for the upcoming {\it TESS} mission,
and the false positive rate is predicted as
30--70\%\cite{2015ApJ...809...77S} depending on the galactic latitude.
Thus one needs to distinguish and exclude false positives caused by eclipsing binaries
to validate the true planetary nature by additional follow-up observations.

To validate the {\it bona fide} planetary nature of each candidate,
we have focused on the capabilities of multi-color transit photometry.
Dimming caused by a true planet should be fairly achromatic,
while that caused by an eclipsing binary would change significantly with wavelength.
We previously developed a multi-color simultaneous camera
named MuSCAT\cite{2015JATIS...1d5001N}
for the 1.88m telescope of National Astronomical Observatory of Japan located in Okayama, Japan,
which is capable of 3-color simultaneous imaging in $g$ (400--550 nm),
$r$ (550--700 nm), and $z_s$ (820--920 nm) bands.
MuSCAT has demonstrated high photometric precisions of less
than 0.05\% as root-mean-square (rms) residuals of 60~s binning
for a 10th magnitude star HAT-P-14\cite{2016ApJ...819...27F}\,\!,
and have successfully validated several transiting
planets\cite{2017PASJ...69...29N,2017Natur.546..514G,2018AJ....155..127H}\,\!.
In the upcoming {\it TESS} era, high-precision multi-color transit photometry
become more important due to the large number of potential candidates of
transiting planets.
To efficiently validate true transiting planets, it is desired to deploy multiple
multi-color simultaneous cameras on 1--2~m class telescopes around the world.
We thus decided to develop the second multi-color simultaneous camera
for the 1.52~m  Telescopio Carlos S\'anchez (TCS) in the Teide Observatory, Canaries, Spain,
which is located at the longitude difference of about 150 deg from
the Okayama observatory.

In addition to the importance for validating true planets,
ground-based instruments for high precision transit photometry
will become more important in the {\it TESS} era due to the following reasons:
First, since the pixel scale of {\it TESS} is 21 arcsec\cite{2015JATIS...1a4003R}
and a typical aperture size is over 1 arcmin, 
there is a high possibility of blending nearby stars
in the same aperture with targets.
For this reason, additional transit observations are necessary
to identify which star is indeed dimming and
to derive precise radii of discovered transiting planets.
Second, the monitoring duration of {\it TESS} for each sector is only about 27 days,
which is significantly shorter than the Kepler (over 4 years) or
K2 (about 80 days) missions.
Thus, it is very important to observe additional transits after the {\it TESS}
observations to improve the transit ephemerides of targets.
This is especially important for selecting potential JWST targets,
as observing times of such space telescopes are
quite valuable\cite{2016AJ....152..171F}\,\!.
Third, for multi-transiting planetary systems in mean motion resonance (MMR)
like TRAPPIST-1\cite{2016Natur.533..221G} or
K2-19\cite{2015A&A...582A..33A,2015ApJ...815...47N}\,\!,
high precision transit observations can determine the masses
of the planets via transit timing variations (TTVs).
TTVs would become a good alternative method to measure the masses of
transiting planets in MMR in the {\it TESS} era.
Finally, optical transit depths of a true planet surrounded by hydrogen dominated atmosphere
have a weak wavelength dependence caused by the nature of the planetary
atmosphere\cite{2013ApJ...770...95F,2016A&A...585A.114P,2018ApJ...853....7K}\,\!.
High precision multi-color transit photometry is useful to probe such weak wavelength
dependence and efficiently select good targets for further follow-up with larger ground-based or
space-based telescopes.

The rest of this paper is organized as follows.
We first describe the TCS 1.52~m telescope in the Teide Observatory (Sec. \ref{sect:TCS}).
We then detail designs of the optical system of MuSCAT2 and characteristics of
its components (Sec. \ref{sect:MuSCAT2}).
We report the performance of MuSCAT2 based on commissioning observations (Sec. \ref{sect:test}).
We finally summarize this paper (Sec. \ref{sect:summary}).

\section{The Carlos S\'anchez Telescope (TCS)}
\label{sect:TCS}

The Carlos S\'anchez telescope
(Latitude: 28$^{\circ}$ 18' 01.8" N, Longitude: 16$^{\circ}$ 30' 39.2" W)
is located at the Teide Observatory (OT), at 2386.75 m over sea level.
The OT is one of the best astronomical observing sites in the world 
with a typical weather success rate of about 80\% and a typical seeing size of about 0.8 arcsec.
TCS has a primary mirror with a diameter of 1.52 meters (60'').
Originally build by ICSTM in collaboration with other groups from the UK and IAC,
it was commissioned in 1971 and started operations in 1972.
It was originally designed as a low cost flux collector for infrared astronomy
and it was one of the earliest telescopes with a thin mirror.
For many years, it was one of the largest telescopes in the world mainly settled
for infrared astronomy.
In 1983 telescope ownership was transferred to the IAC and major improvements
in order to keep the telescope in a competitive status were performed,
including automating control and data acquisition, development of instrumentation
for common use and improving the pointing precision and tracking. 

The TCS main mirror is fixed in an equatorial structure with a Cassegrain focus
and focal length of f/13.8 in a Dall-Kirkham type configuration.
Its common user-instrumentation includes two other instruments:
CAIN and FastCam.
More technical details on the TCS and current instruments can be found at:\\
\verb|http://www.iac.es/OOCC/instrumentation/telescopio-carlos-sanchez/|

\section{The MuSCAT2 Instrument}
\label{sect:MuSCAT2}  

\subsection{Instrumental Optical Design}
\label{subsec:design}

   \begin{figure}
   \begin{center}
   \begin{tabular}{c}
   \includegraphics[width=9cm, angle=-90]{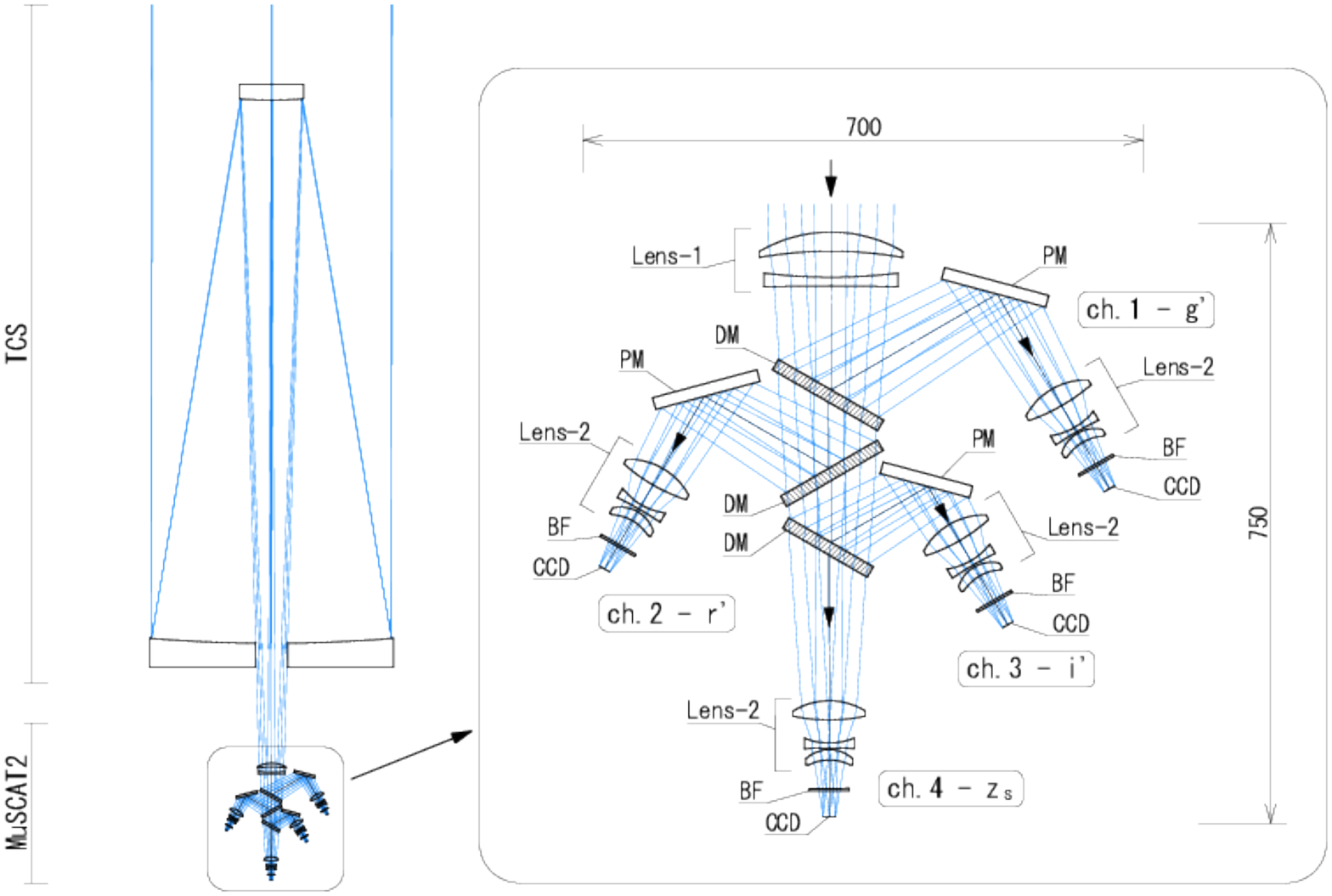}  
   \end{tabular}
   \end{center}
   \caption 
   { \label{layout} 
Layout of the optical system of TCS and MuSCAT2.
The left side shows ray trace for the whole telescope and instrument system.
The right side is an enlarged view of ray trace for MuSCAT2.
Incident light is separated into 4 channels by 3 dichroic mirrors (DMs).
F-conversion lenses are placed before (Lens-1) and after (Lens-2) the DMs.
Plane mirrors (PMs) are inserted in the channels 1-3 to fold optical path. 
Bandpass filters (BFs) are inserted just before the CCD cameras.
Instrument sizes are described in mm.
} 
%
   \begin{center}
   \begin{tabular}{c}
   \includegraphics[width=8cm, angle=-90]{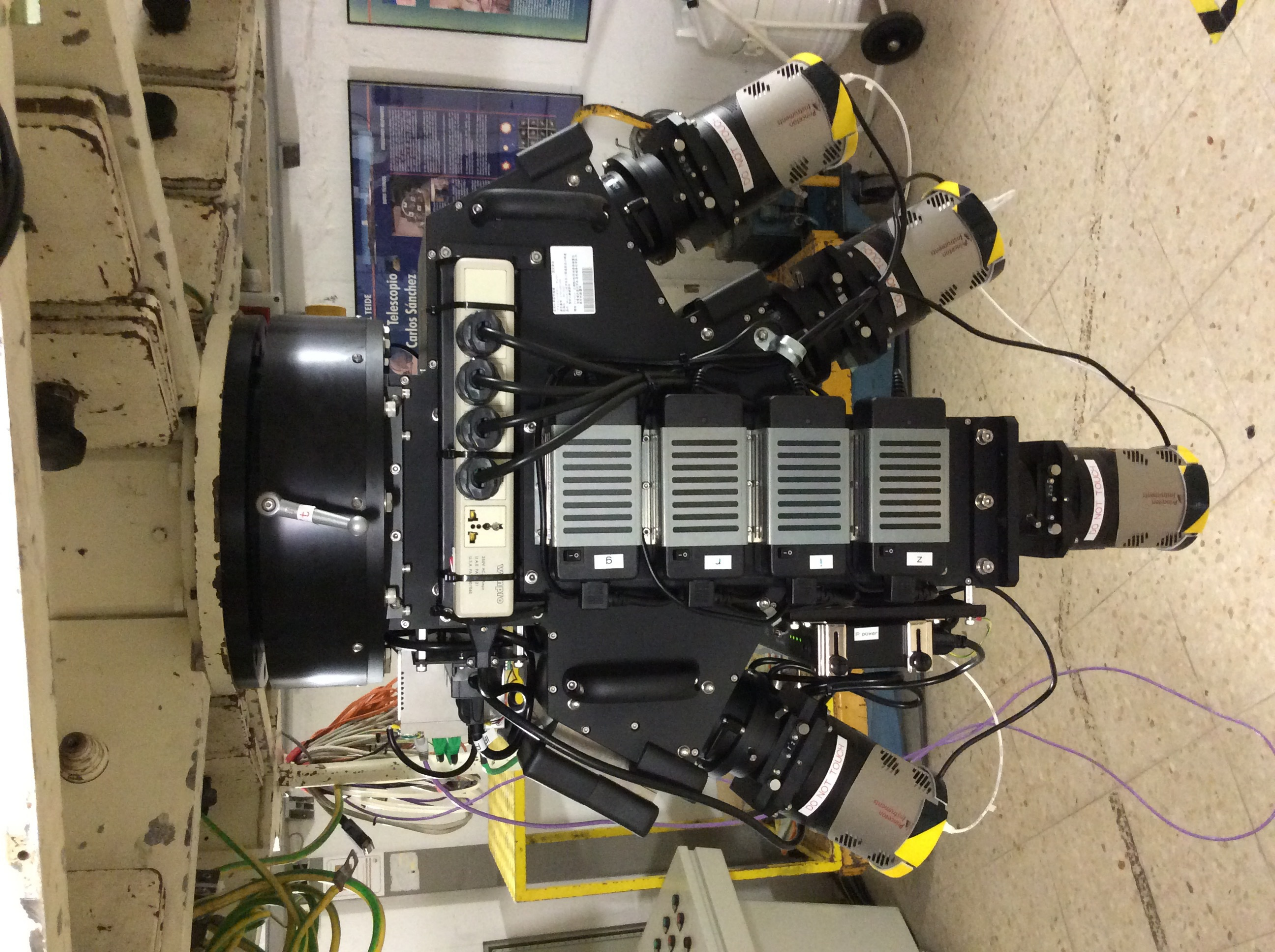}  
   \end{tabular}
   \end{center}
   \caption 
   { \label{picture} 
A picture of MuSCAT2 installed at the Cassegrain focus of the TCS 1.52~m telescope.
The orientation of the picture corresponds to the right side of Fig.~\ref{layout}.
} 
   \end{figure} 

The optical layout of MuSCAT2 is shown in Figure~\ref{layout}.
MuSCAT2 is installed at the Cassegrain focus of the TCS as pictured in Figure~\ref{picture}.
The optical system of MuSCAT2 is composed of F-conversion lenses
(Lens-1 and Lens-2) to widen the field-of-view (FoV) and 
dichroic mirrors (DMs) to separate light into 4 wavelength bands.
The F-conversion lenses are placed before and after the DMs, and play
roles to make F-number faster (focal image brighter) and to correct
off-axis coma aberration characteristic of a Dall-Kirkham type telescope.
The Lens-1 roughly correct off-axis coma aberration of all bands,
and convert F-number from f/13.8 to f/8.
The Lens-2, which are not identical but optimized for each band,
further correct coma aberration of each band,
and convert F-number from f/8 to f/4.2.
All the lenses are applied anti-reflection (AR) coating of less than 1\%
reflectance ratio.
A secondary mirror position is about 34 mm closer from the nominal position
to obtain focused images on the detectors of MuSCAT2.

\begin{figure}
   \begin{center}
   \begin{tabular}{c}
   \includegraphics[width=9cm]{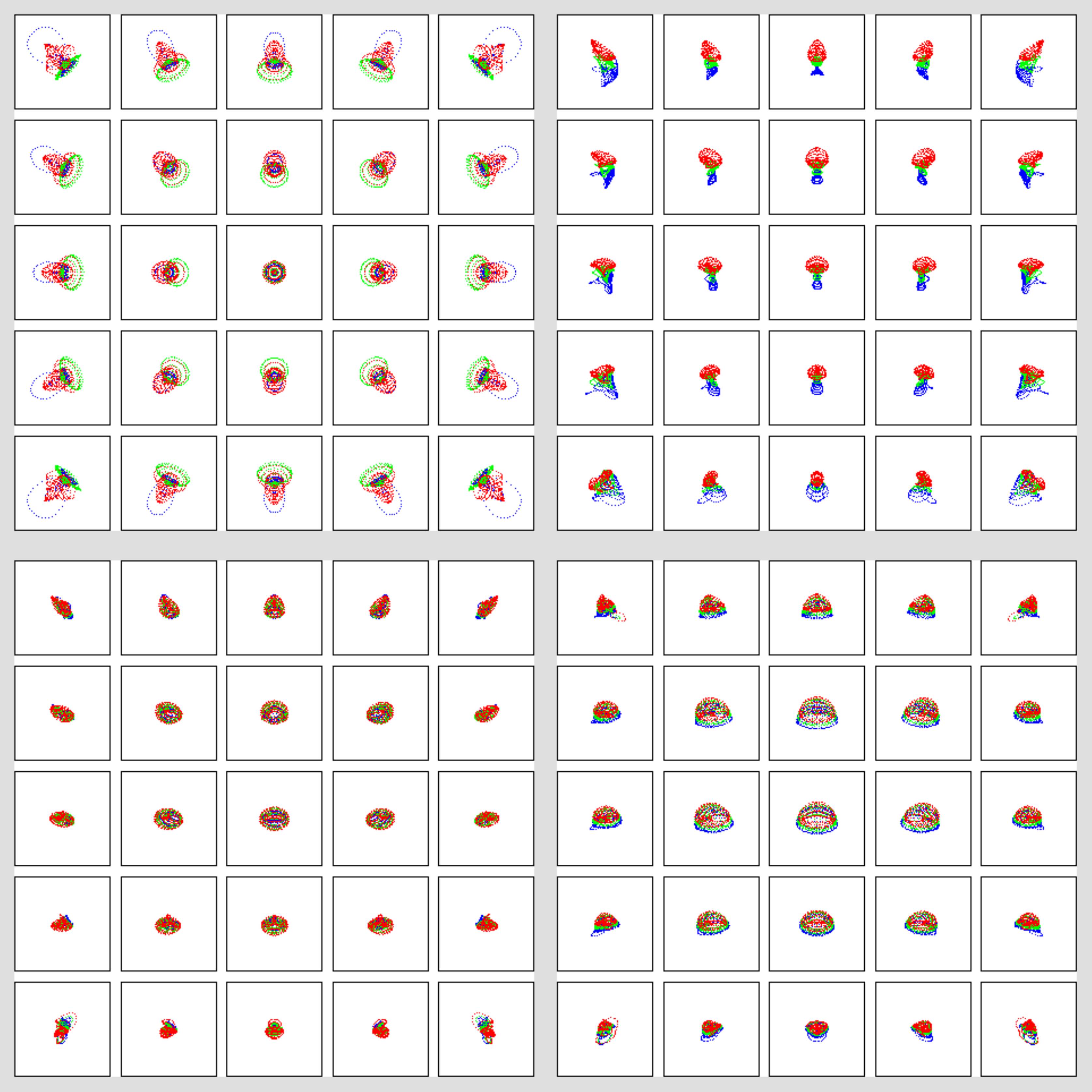}  
   \end{tabular}
   \end{center}
   \caption 
   { \label{spotfocus} 
Spot diagram for on-focus cases for ch.~1 (upper left), ch.~2 (upper right),
ch.~3 (lower left), and ch.~4 (lower right).
The 5$\times$5 cells represent the FOV of 1k$\times$1k CCD.
The size of each cell corresponds to 1 arcsec.
Colors indicate simulated images of the shortest (blue), mid (green), and longest (red)
wavelength in each channel.
Specifically, 400, 470, 550 nm for ch.~1, 
550, 630, 700 nm for ch.~2,
700, 760, 820 nm for ch.~3, and
700, 800, 950 nm for ch.~4.
} 
%
   \begin{center}
   \begin{tabular}{c}
   \includegraphics[width=9cm]{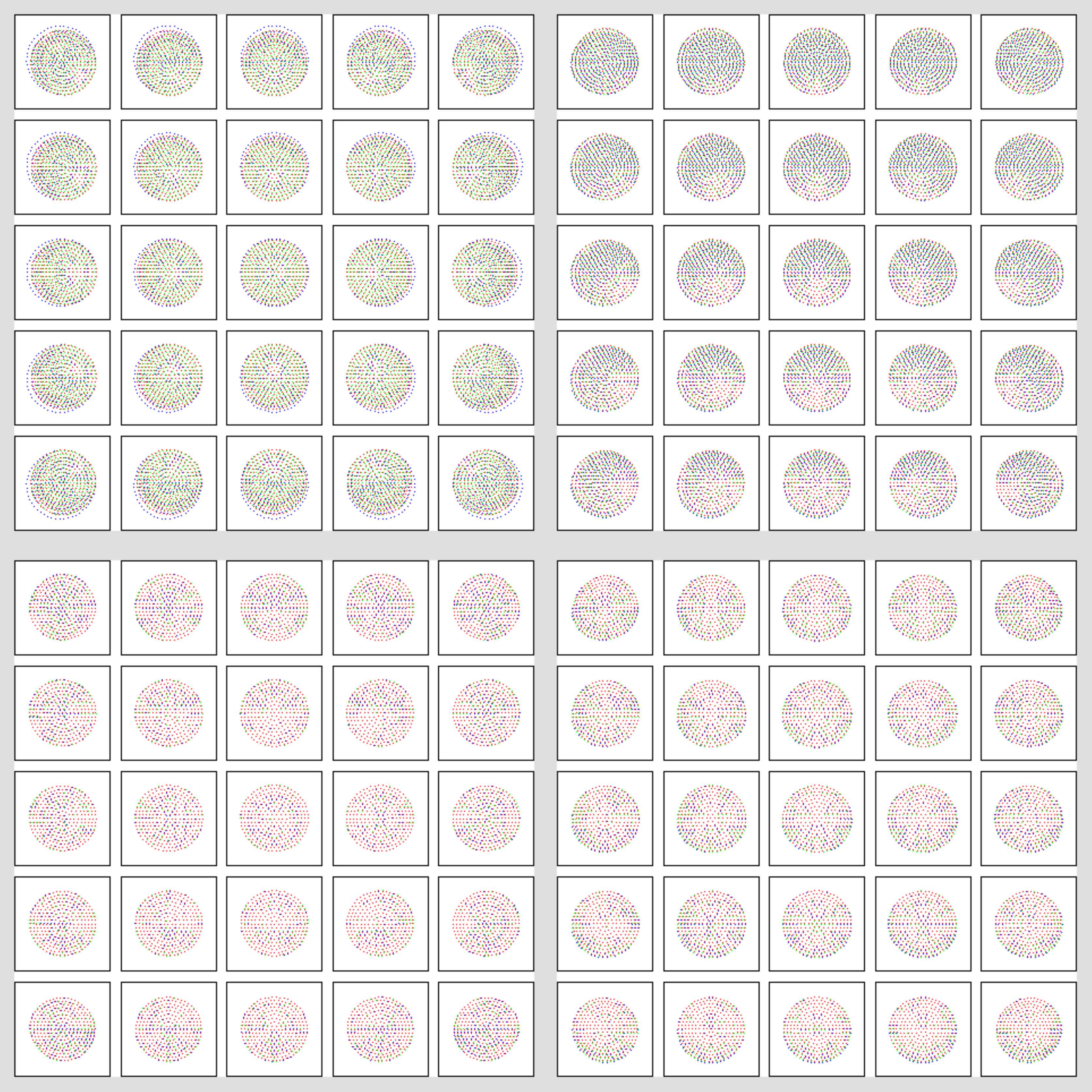}  
   \end{tabular}
   \end{center}
   \caption 
   { \label{spotdefocus} 
Same as Figure~\ref{spotfocus}, but for defocused cases where
the secondary mirror is moved by 5~mm from the focused position.
In this case, the size of each cell corresponds to 10 arcsec.
} 
   \end{figure}

Simulated spot diagrams on CCD cameras of MuSCAT2 are shown in
Figures \ref{spotfocus} and \ref{spotdefocus}.
Figure~\ref{spotfocus} plots spot diagrams for on-focus cases.
The figure indicates spot radius of all wavelength bands are
well smaller than 0.5 arcsec throughout the FoV.
This means the imaging performance of MuSCAT2 is sufficiently
good compared to the typical seeing (0.8 arcsec) of the Teide observatory.
Figure~\ref{spotdefocus} does for defocused cases where
the secondary mirror is shifted by 5~mm from the focused position,
which makes the spot radius expand to 5-6 arcsec.
The figure implies that images are nearly circular throughout the FoV
and suitable for aperture photometry.

\subsection{Dichroic Mirrors}
\label{DM}

\begin{figure}
   \begin{center}
   \begin{tabular}{c}
   \includegraphics[width=10cm]{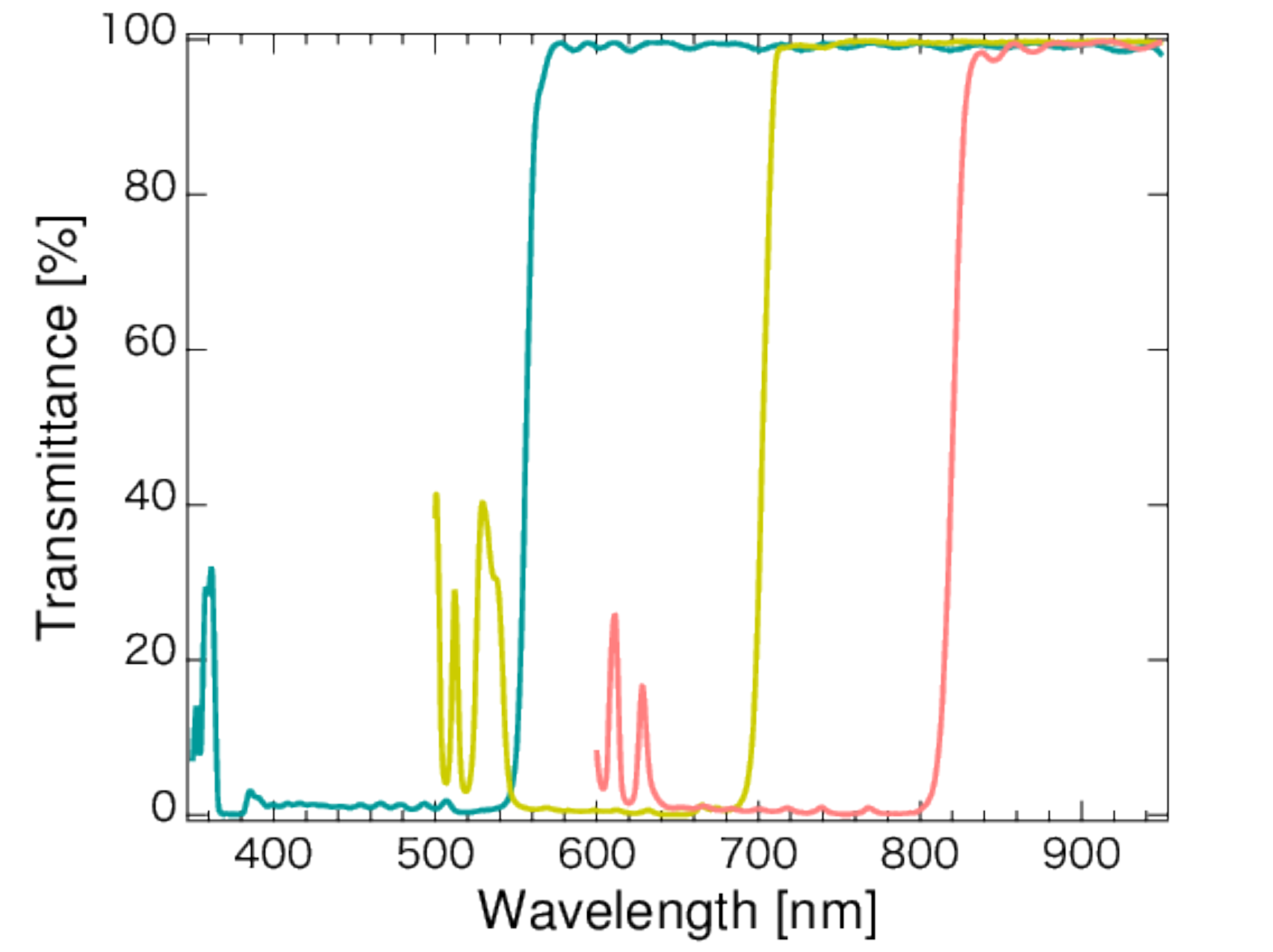}  
   \end{tabular}
   \end{center}
   \caption 
   { \label{DMreflectance} 
Transmittance of the DM1 (left, emerald green), DM2 (middle, yellow), and DM3 (right, pink).
} 
   \end{figure}

Three dichroic mirrors (DMs) to separate incoming light into 4 wavelength channels are
manufactured by Asahi Spectra Co.,Ltd.
The sizes and wedge angles of the DMs are summarized in Table~\ref{DMsummary}.
The reflectance of the DMs is shown in Figure~\ref{DMreflectance} and
the DMs transmit remaining lights almost completely due to
anti-reflection coating  processed on the back sides of the DMs.
All the DMs are inserted with an incident angle of 30 deg.
Plain mirrors are placed after DMs for ch.~1, 2, and 3 to optimize
mechanical structure.

\begin{table}
\caption{
\label{DMsummary}
Sizes and wedge angles of the dichroic mirrors.
}
\begin{center}
\begin{tabular}{c|cccc}
\hline \hline
 & height (mm) & width (mm) & depth (mm) & wedge angle (arcmin) \\
\hline
DM1 & 151.87 & 141.87 & 15.37 & 8 \\
DM2 & 139.87 & 131.90 & 15.37 & 9 \\
DM3 & 121.88 & 119.89 & 15.38 & 10.9 \\
\hline
\end{tabular}
\end{center}
\end{table}

   \begin{figure}
   \begin{center}
   \begin{tabular}{c}
   \includegraphics[width=10cm]{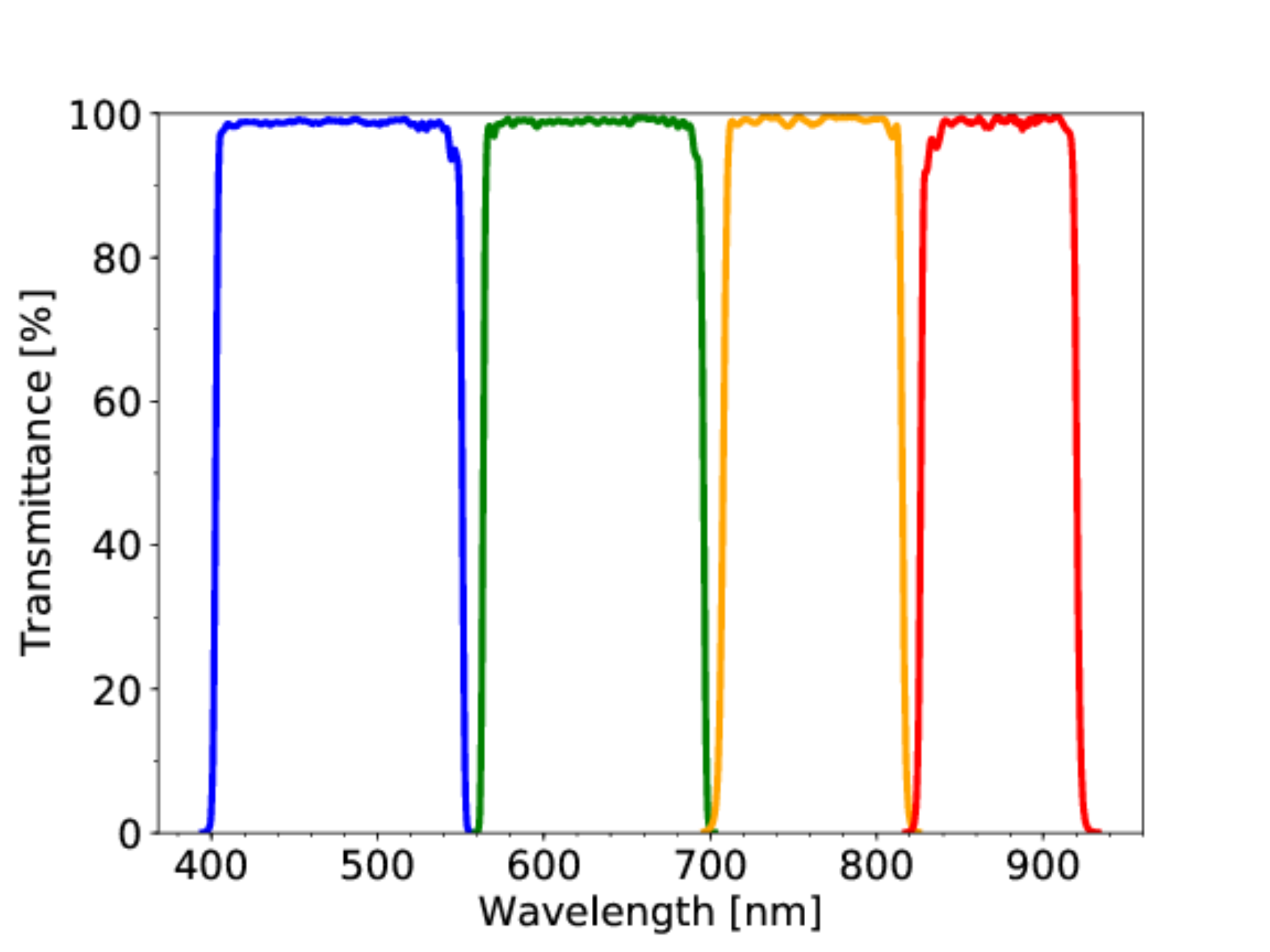}
   \end{tabular}
   \end{center}
   \caption 
   { \label{filterproperty}
Transmittance of $g$ (blue), $r$ (green), $i$ (orange), and $z_{s}$ (red) bandpass filters from left to right.
The gaps of the bandpass filters are coincided with the rises of transmittance shown in
Fig.~\ref{DMreflectance}.
} 
   \end{figure} 

   \begin{figure}
   \begin{center}
   \begin{tabular}{c}
   \includegraphics[width=10cm]{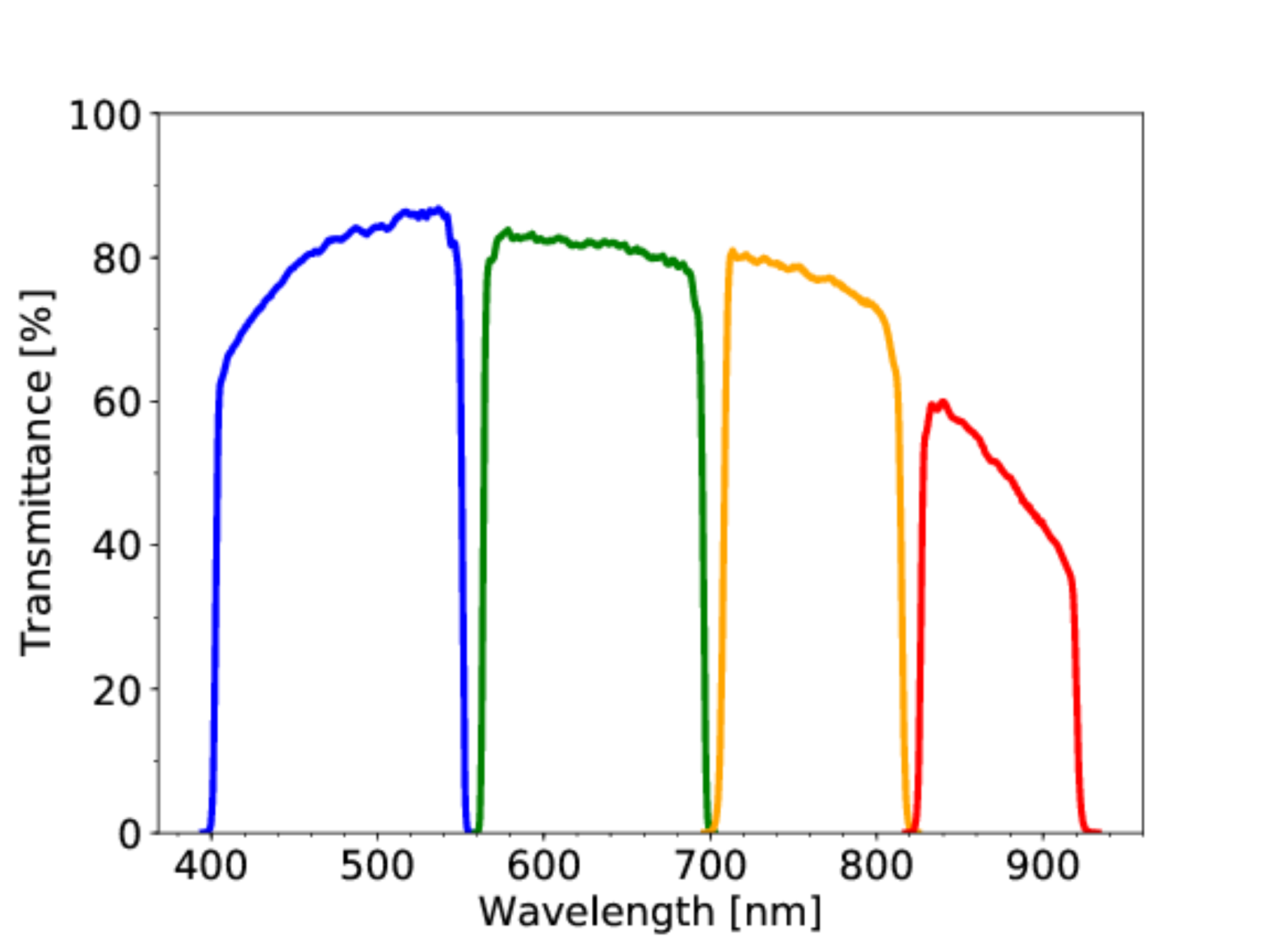}
   \end{tabular}
   \end{center}
   \caption 
   { \label{transmittance_inst}
Total transmittance of the MuSCAT2 instrument in $g$ (blue), $r$ (green), $i$ (orange), and $z_{s}$ (red) bands from left to right.} 
   \end{figure} 

\subsection{Bandpass Filters}
\label{BF}

Following the experience on MuSCAT,
we adopt commercially-available $g'$, $r'$, and $z_{s}$ band filters
of Astrodon Photometrics Generation 2 Sloan filters for ch.~1, 2, and 4, respectively.
The three filters are the same with those for MuSCAT on Okayama 1.88m telescope.
On the other hand, since the $i'$ band filter of Astrodon Photometrics Generation 2
Sloan filters has an unavoidable overlap of transparent wavelength with
the $r'$ and $z_{s}$ band filters, we adopt a custom-ordered $i$ band filter
manufactured by Asahi Spectra Co.,Ltd.
For simplicity, hereafter we call those channels as $g$, $r$, $i$, and $z_s$ bands, respectively.
We note that the reflectance and transmittance of the DMs are optimized for
those 4 bandpass filters.
The size of those filters are 50 mm by 50 mm.
Figure \ref{filterproperty} plots the measured transmittance of the bandpass filters.

\subsection{CCD Cameras}

MuSCAT2 equips 4 CCD cameras manufactured by Princeton Instruments:
one is PIXIS:~1024B used for the ch~2 ($r$ band) and
the others are PIXIS:~1024B\_eXcelon used for the other 3 channels
($g$, $i$, and $z_{s}$ bands).
Each PIXIS camera equips an e2v 47-10 AIMO (Advanced Inverted Mode Operation) science grade 1 CCD.
The cosmetic quality of the CCD is very high.
The CCD specification declares there are less than 200 bad pixels among
$10^6$ pixels and there is no column defect.
Basic specifications (e.g., size, weight, etc) of those CCD cameras are identical with MuSCAT on
the Okayama 188cm telescope (Table~1 of the paper for MuSCAT \cite{2015JATIS...1d5001N})\,\!.
Individual characteristics such as values of dark current and readout noise are summarized in
Table~\ref{CCDsummary}, measured by the manufacturer as inspection verification.
Total transmittance of the MuSCAT2 instrument, including all the optics and the quantum efficiency
of the detectors, is plotted in Figure~\ref{transmittance_inst}.
Each CCD camera is independently controllable from an instrument operating PC so that
observers can set either individually-different exposure times or synchronized ones,
depending on science cases.
Typical time interval between two exposures due to readout and fits file creation is 1-4 s.
Engineering test results for (non-)linearity and saturation level of the individual CCDs are shown
in Section~\ref{subsec:nonlinearity}.

\begin{table}
\caption{
\label{CCDsummary}
Dark current and readout noise of CCDs based on inspection verification by the manufacturer.
}
\begin{center}
\begin{tabular}{c|ccccc}
\hline \hline
CCD & band & dark current & readout noise (100 kHz) & readout noise (2 MHz) & gain \\
& & (e$^{-}$ pix$^{-1}$ s$^{-1}$) & (e$^{-}$ pix$^{-1}$ rms) & (e$^{-}$  pix$^{-1}$ rms) & (e$^{-}$ ADU$^{-1}$) \\
\hline
ch.1 & $g$  & 0.0012  & 4.08 & 12.35 & 1.04\\
ch.2 & $r$  & 0.0003  & 3.96 & 11.51 & 0.96 \\
ch.3 & $i$  & 0.00047 & 4.02 & 13.13 & 1.03 \\
ch.4 & $z_s$ &  0.00047 & 4.48 & 12.56 & 1.02 \\
\hline
\end{tabular}
\end{center}
\end{table}

\subsection{Instrument Rotator}

As the TCS is not equipped with an instrument rotator at the Cassegrain focus,
we installed a custom-made instrument rotator between the telescope flange and MuSCAT2.
The reason why the instrument rotator is attached is that it is important to obtain
good comparison stars in the field of view to enable high precision transit photometry.
The presence of the instrument rotator makes it possible to obtain a potential comparison
star located at 1.4 times farther than the length of the side of the field of view.
The rotator was designed and manufactured by a limited company, CHUO-OPT, Japan.
The rotator is controllable through a serial cable from a Linux-based PC that is attached
to MuSCAT2 (the PC also controls the CCD cameras).
While the angular resolution of the rotator is 0.89'', the absolute accuracy of pointing
a certain position angle is about 0.1 degrees, which is limited by the inaccuracy of
re-mounting of the rotator to the telescope (together with MuSCAT2 to exchange instruments)
and that of determining origin using a photomicro sensor.
The range of the angle that the rotator can mechanically rotate is $\pm$90 degrees,
however, it is currently operated in the range of $\pm$45 degrees for safety reason.
Although the gear train has basically no backlash, it can move a little bit depending on
the telescope position due to mechanical flexure, which alters the position of stars
on the detectors by up to $\sim$1".

\section{Commissioning Observations and Results}
\label{sect:test}  

First-light commissioning of MuSCAT2 was made on August 24th (the night of August 23th), 2017 UT.
Subsequently, we conducted commissioning observations through 2017 and early 2018.
We have examined the performance of MuSCAT2 on the TCS 1.52~m telescope.
We summarize the results of the commissioning observations below.

\subsection{Field of View}

We have derived pixel scales and corresponding field of view (FoV) of MuSCAT2
based on images of NGC6885 using TOPCAT\cite{2005ASPC..347...29T} and ccmap in IRAF.
The derived pixel scales and FoV are
0.44 arcsec per pixel and 7.4$\times$7.4 arcmin$^2$ for all the bands.
Thanks to the presence of the instrument rotator, it is possible to obtain a potential comparison
star located at about 10 arcmin away from a target star.
We have confirmed that centers of FoV of 4 CCD cameras are adjusted
within 10 pixels in both x and y directions,
and relative rotations of the position angle fit within 1 deg.

\subsection{Sensitivity and Efficiency}

   \begin{figure}
   \begin{center}
   \begin{tabular}{c}
   \includegraphics[width=12cm]{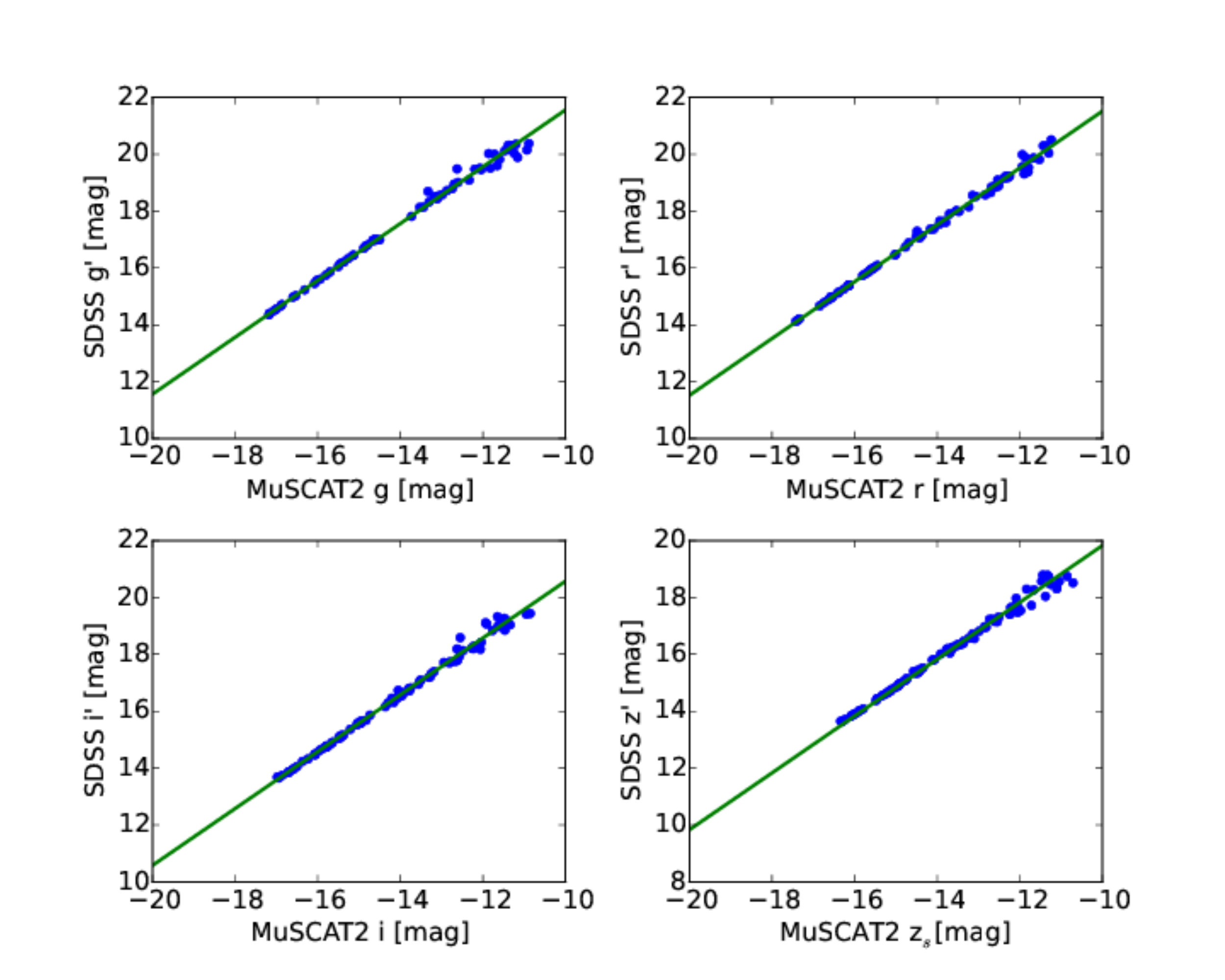}
   \end{tabular}
   \end{center}
   \caption 
   { \label{zero}
Relations between MuSCAT2 instrumental magnitudes (horizontal) and
the SDSS catalog magnitudes (vertical) of the stars for $g$ (upper-left), $r$ (upper-right),
$i$ (lower-left), and $z_s$ (lower-right) bands, respectively.
Green solid lines represent the best-fit linear functions.
Zero point magnitudes are derived as y-intercept of the best-fit functions.} 
%
\vspace{-5mm}
   \begin{center}
   \begin{tabular}{c}
   \includegraphics[width=12cm]{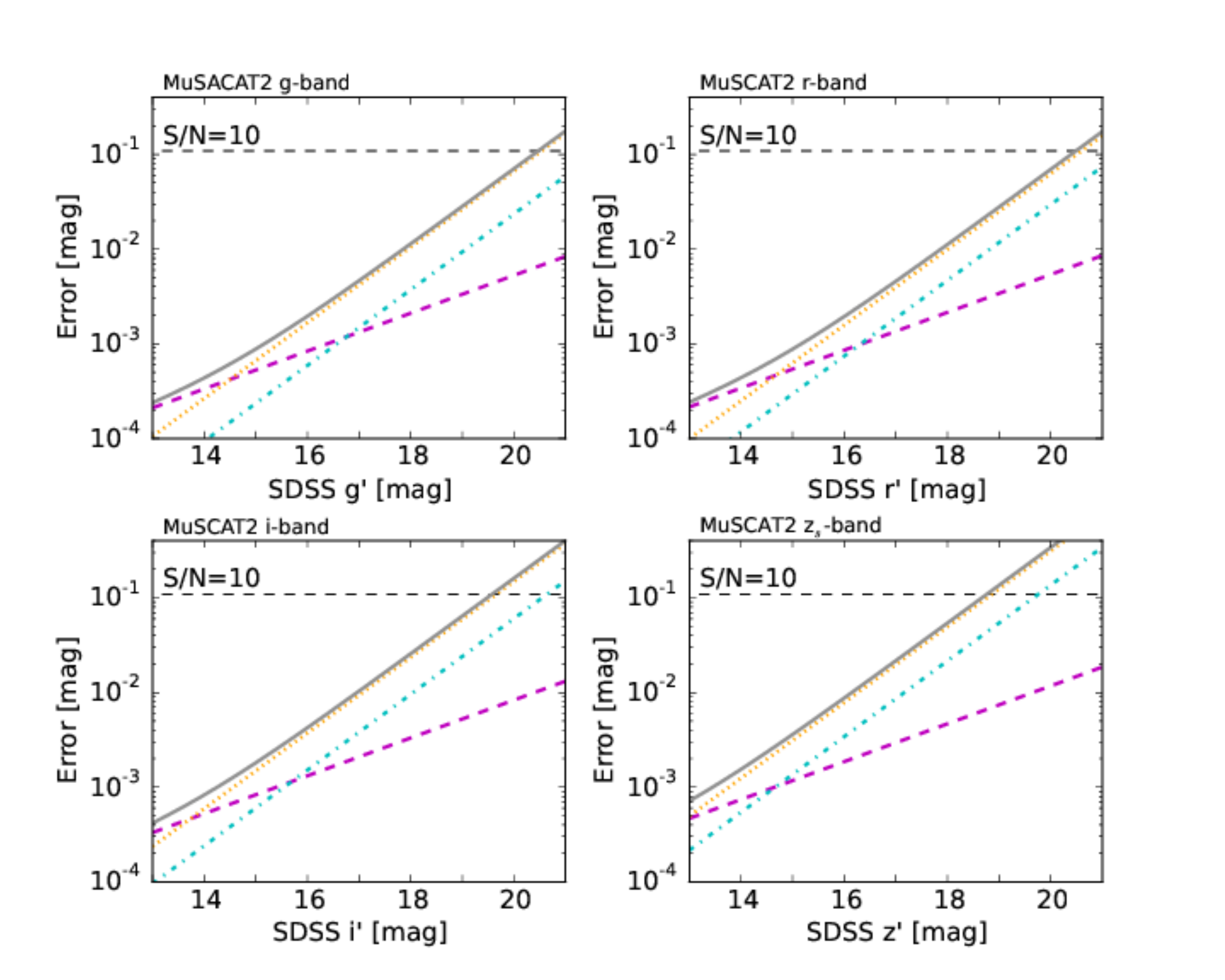}
   \end{tabular}
   \end{center}
   \caption 
   { \label{limitingmag}
Relations of SDSS magnitudes (horizontal) and expected errors
for 600 s
in magnitude (vertical) for $g$ (upper-left), $r$ (upper-right), $i$ (lower-left),
and $z_s$ (lower-right) bands, respectively.
Black solid, magenta dashed, orange dotted, and aqua dot-dashed lines represent
total noise, stellar noise, readout noise, and sky noise, respectively.
} 
   \end{figure} 

\begin{table}
\begin{center}
{\footnotesize
\caption{Summary of throughput (TP) of MuSCAT2 on the TCS 1.52m telescope.}
\label{throughput}
\scalebox{0.9}{
\begin{tabular}{ccccccc}\hline
band& sky transmittance & M1 & M2 & MuSCAT2 (total) & expected TP & measured TP\\ \hline
$g$ & 81\%$^a$ & 70\%$^b$ & 70\%$^b$ & 80\%&  32\% & 26\% \\
$r$ & 89\%$^a$ & 70\%$^b$ & 70\%$^b$ & 81\%& 35\% & 33\% \\
$i$ & 92\%$^a$ & 70\%$^b$ & 70\%$^b$ & 77\%& 35\% & 22\% \\
$z_{s}$ & 95\%$^a$ & 70\%$^b$ & 70\%$^b$ &  51\% & 24\% & 15\% \\ \hline
\end{tabular}
}}
{\scriptsize \\ M1 = primary mirror, M2 = secondary mirror, MuSCAT2 (total) = all optics including filters, CCD QE, and BBAR coating,\\
$^a$: based on a model atmosphere by the libRadtran software package\cite{2005ACP.....5.1855M}\,\!, $^b$: assumed}
\end{center}
\end{table}

We have estimated zero point magnitudes of MuSCAT2 using the images of M67 as follows.
For each band, 300 frames with 2 s exposure were obtained,
but 7 frames were discarded because they had split PSF due to
large guiding feedback during exposure.
Remaining 293 frames were bias-subtracted, flat-fielded, position-aligned,
and combined into a master image in each band.
We then employed SExtractor\cite{1996A&AS..117..393B}\,\!, ccmap in IRAF,
and TOPCAT\cite{2005ASPC..347...29T}
to identify stars in the master images and to match them with
the SDSS DR7 catalog\cite{2009ApJS..182..543A}\,\!.
As a result, we matched about 80 stars in each band.
We measured fluxes of the stars and sky background in each band from the master image,
and multiplied by 300/293 to make them correspond to 600 s exposure.
The fluxes of the stars and sky background were then converted into magnitudes.
The magnitudes of sky background were
$g' = 20.4$ mag arcsec$^{-2}$, $r' = 19.8$ mag arcsec$^{-2}$,
$i' = 19.0$ mag arcsec$^{-2}$, and $z' = 18.2$ mag arcsec$^{-2}$, respectively.
We fitted relations between MuSCAT2 instrumental magnitudes and the SDSS catalog magnitudes
of the stars in each band by a linear function, and derived zero point magnitudes
(corresponding to 1 ADU per 600 s)
of MuSCAT2 as $g' = 31.56$ mag, $r' = 31.52$ mag, $i' = 30.58$ mag, and $z' = 29.83$ mag, respectively
(see Figure~\ref{zero}).
We note that we neglect color terms and simply approximate that
the $g$, $r$, $i$, and $z_{s}$ bands of MuSCAT2 are identical to
the SDSS $g'$, $r'$, $i'$, and $z'$ bands.

We then calculated ratios of the stellar noise ($n_{\rm stellar}$), the sky noise ($n_{\rm sky}$),
the readout noise ($n_{\rm read}$) to the signal using the following equations:
\begin{eqnarray}
n_{\rm stellar} / {\rm signal} = (10^{-0.4(m-Z)})^{0.5} \,\, / \,\, (10^{-0.4(m-Z)}),\\
n_{\rm read} / {\rm signal} = {n \, (A \, N_{\rm exp})^{0.5}} \,\, / \,\, (10^{-0.4(m-Z)}),\\
n_{\rm sky} / {\rm signal} = (A\,10^{-0.4(sky-Z)})^{0.5} \,\, / \,\, (10^{-0.4(m-Z)}),
\end{eqnarray}
where $m$ is the MuSCAT2 instrumental magnitude,
$Z$ is the zero point magnitude,
$n$ is the readout noise in ADU per pixel (see Table~\ref{CCDsummary}),
$A$ is the area of the aperture (a circle with the radius of 1.7 times of FWHM),
$N_{\rm exp}$ is the number of exposures, and
$sky$ is the sky background magnitudes.
We compute square sum of those noises (total noises), and plot them in Figure~\ref{limitingmag}.
We derive limiting magnitudes of MuSCAT2 giving the signal-to-noise (S/N) ratio of 10
with 10-min exposure as
$g'_\mathrm{lim} = 20.5$ mag, $r'_\mathrm{lim} = 20.5$ mag,
$i'_\mathrm{lim} = 19.7$ mag, and $z'_{\mathrm{lim}} = 19.0$ mag.

Finally, we estimate total throughput (TP) of the TCS 1.52m telescope and MuSCAT2 by
comparing measured fluxes coming from stars with expected fluxes from known magnitudes.
The measured total throughput values are 
26\%,  33\%,  22\%, and 15\%, respectively for in $g$, $r$, $i$, and $z_{s}$ bands.
We also estimate expected total throughput as shown in Table~\ref{throughput},
considering transmittance of sky, reflectivity of the primary and secondary mirrors,
transmittance and reflectivity of MuSCAT2 optics (including lens, mirror, DMs, filters),
and quantum efficiency (QE) of detectors with broadband antireflection (BBAR) coating on the camera window.
To calculate the sky transmittance, we employ the libRadtran software
package\cite{2005ACP.....5.1855M} to model the telluric atmosphere with aerosol
at the elevation of 2390 m.
Since the aerosol model is not definitive but has large uncertainty,
the expected TP has uncertainties of several percent.
We also note that we neglect the effect of spiders, which would cause additional a few percent loss of light. 
Consequently, we found that the TP in $g$ and $r$ bands are roughly in agreement with the expected ones,
but that in $i$ band is a bit lower and that in $z_{s}$ band is significantly worse than expected.
Although an exact reason is uncertain, we suspect that this is due to degradation of reflectivity of
the primary and secondary mirrors due to micron-sized dusts on surfaces of them.

\subsection{CCD Non-linearity}
\label{subsec:nonlinearity}

We have examined the linearity and saturation level of the individual CCDs
by taking dome flat images with various exposure times.
The illumination source of the dome flat is composed of two types of lamps,
one is fluorescent lamp and the other is voltage-controllable filament lamp,
both distributed on the wall.
Because the illuminance of these lamps is not stable but time variable,
we have taken the following procedure to mitigate this effect.
First, for each CCD, we adjusted the voltage of the filament lamps so that
the average count of a flat image with the exposure time of 10~s became about 30,000~ADU.
Second, we obtained a set of dome-flat images with a range of exposure times
from 1~s through 23~s, with which most of the pixels are saturated, being incremented by 1~s.
In addition to those ``experimental'' data, we also took ``reference'' images with
the exposure time of 10~s in between two consecutive experimental exposures.
We repeated this set of exposures three times for each CCD.

\begin{figure}
   \begin{center}
   \begin{tabular}{c}
   \includegraphics[width=11cm]{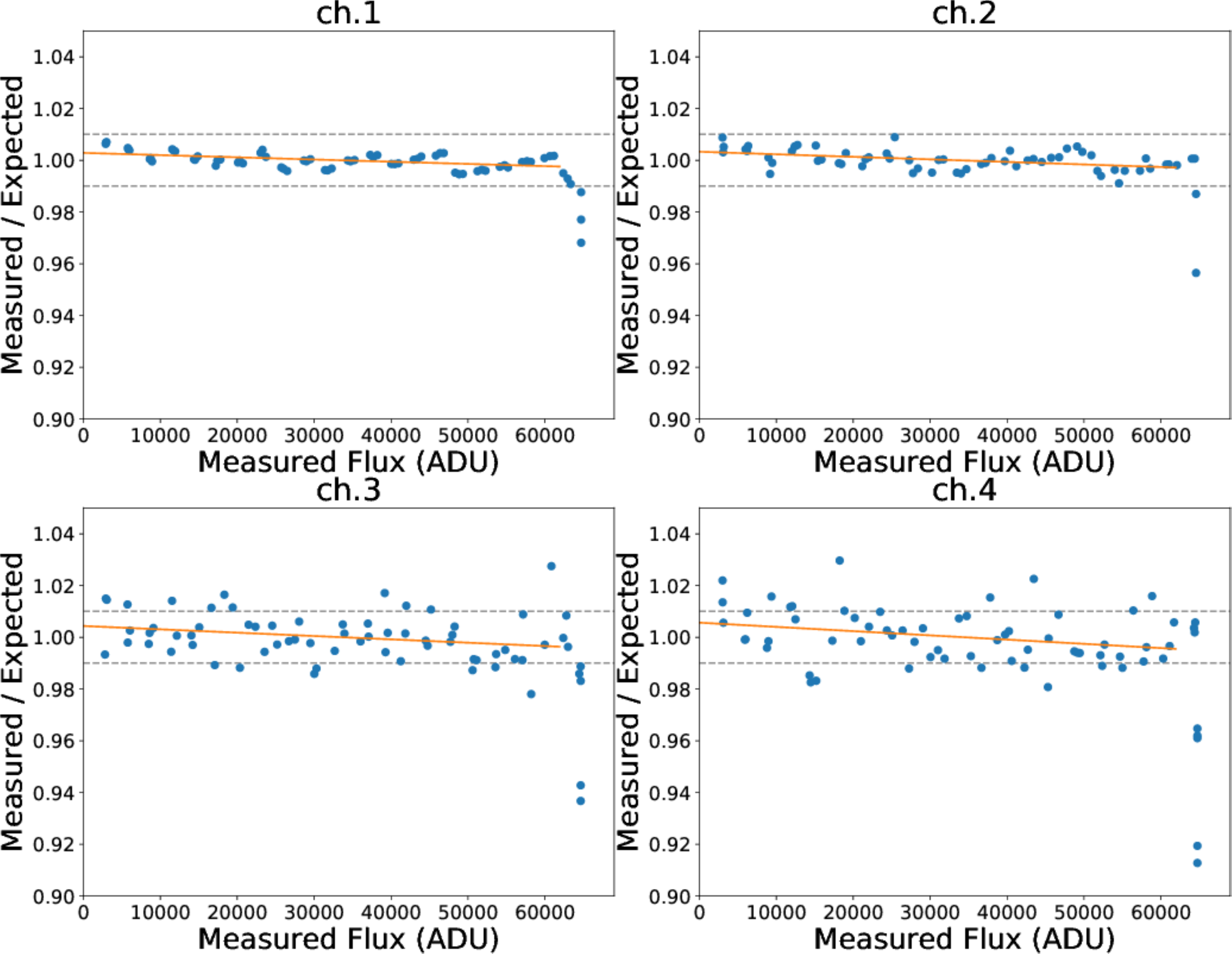}  
   \end{tabular}
   \caption 
   { \label{ccdnonlinear} 
Relations of measured fluxes (horizontal) and ratios of measured fluxes to expected ones (vertical)
for CCD cameras for $g$ (upper-left), $r$ (upper-right), $i$ (lower-left), and $z_s$ (lower-right) bands, respectively.
The orange solid lines represent the best-fit linear function models.
The gray dotted lines indicate boundaries of 1\% non-linearity. 
} 
\includegraphics[width=8cm]{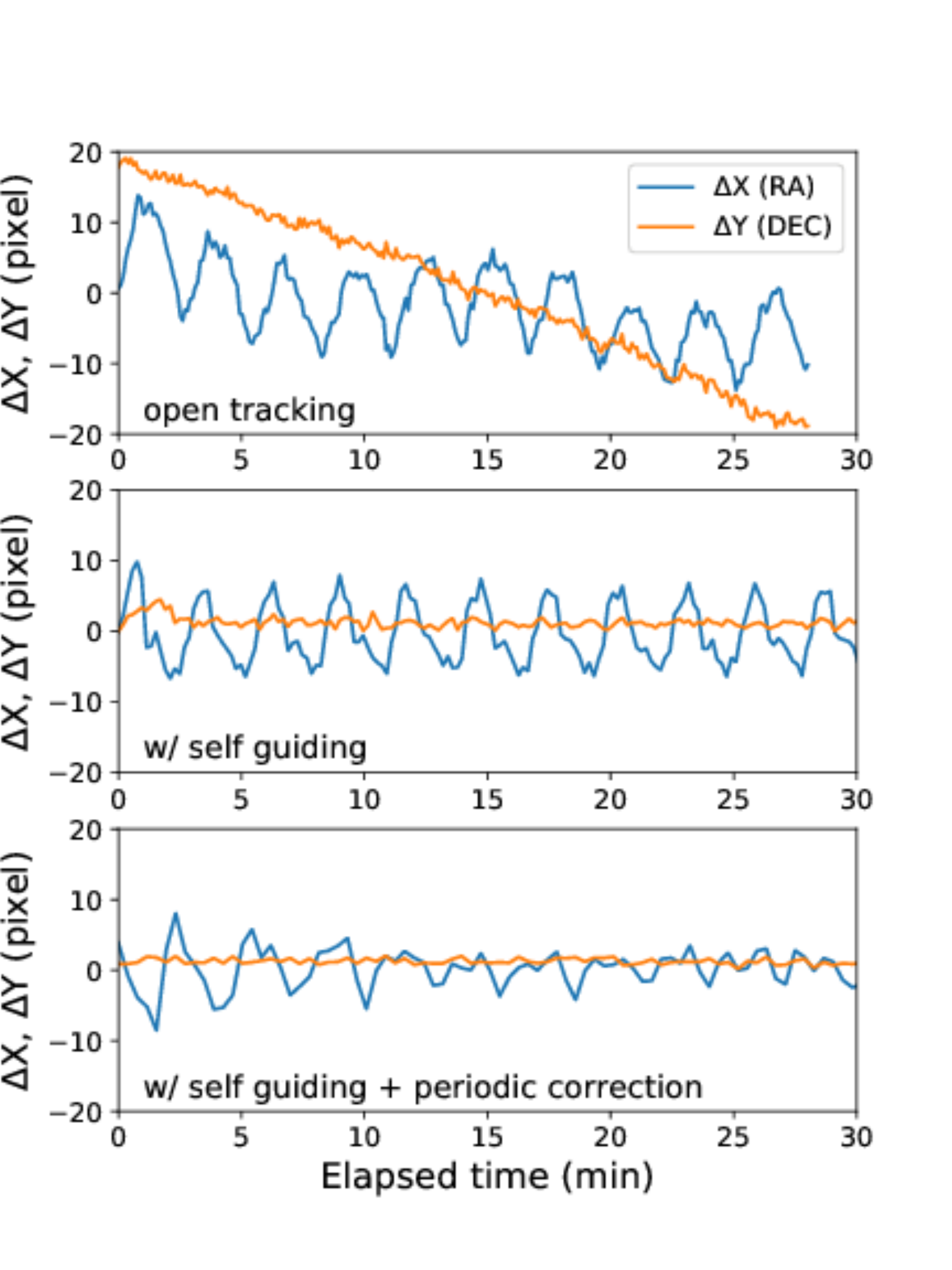}
\caption{Sample data of stellar positions on a CCD as a function of time.
The X (RA) and Y (DEC) positions are shown in orange and blue, respectively.
(Top) the data taken without any guiding and correction.
(Middle) the data taken with the self auto guiding activated.
(Bottom) the data taken with both the self guiding and periodic correction for tracking applied.
Note that these three datasets were taken on different nights for different targets.
The declinations of the targets were 0, 0, and -11 degrees
for the top, middle, and bottom datasets, respectively.
\label{fig:dxdy}}
\end{center}
\end{figure}

After subtracting a dark image from all of the exposed images,
we have calculated average counts on the individual exposed images.
Then, for each experimental image we have estimated the expected count
by multiplying the average count of the adjacent reference images by $t_\mathrm{exp}/10$,
where $t_\mathrm{exp}$ is the exposure time of an experimental image in seconds.
In Figure~\ref{ccdnonlinear}, we show the ratio of the measured average count to
the expected one calculated for the experimental images as a function of
the measured count for the respective CCDs.
In all plots, the measured-to-expected ratio is distributed around unity up to 62,000-64,000 ADU,
above which the value drops due to saturation.
Below the saturation level, the measured-to-expected ratio gradually decreases with the measured count.
We fit the data below 62,000 ADU with a linear function, which gives the following relations:
\begin{eqnarray}
y &=& -0.00085  x + 1.00283\ \mathrm{(ch.1)},\\
y &=& -0.00098  x + 1.00328\ \mathrm{(ch.2)},\\
y &=& -0.00129  x + 1.00435\ \mathrm{(ch.3)},\\
y &=& -0.00164  x +  1.00563\ \mathrm{(ch.4)},
\end{eqnarray}
where $x$ and $y$ are the measured count $\times10^{-4}$ (ADU) and
the measured-to-expected ratio, respectively.
The best-fit models are plotted by an orange solid line in Figure~\ref{ccdnonlinear}.
These results indicate that the non-linearity of all the CCDs are
well below 1\% up to $\sim$62,000 ADU.
According to Mann et al. (2011)\cite{2011PASP..123.1273M}\,\!,
the effect of CCD non-linearity on photometric precision can be written as,
rms $\sim\sqrt{2} \alpha \sigma^2 / \sqrt{N}$,
where $\alpha$ is the level of non-linearity in the pertinent ADU range, $\sigma$ is
the standard deviation of normalized incident flux, and $N$ is the number of pixels in an aperture of interest.
In our observation, $N$ is typically 300-1000.
If we assume α=0.01 (the non-linearity level of 1\%) and a conservative atmospheric condition of σ=0.3
(incident flux varies by 30\% rms), the photometric noise arising from the CCD non-linearity is less than 0.008\%.
Therefore, the CCD non-linearity up to ~62,000 ADU is basically negligible for transit photometry with MuSCAT2.

\subsection{Self Auto-guiding and Correction of Periodic Tracking Error}
\label{subsec:guiding}

Tracking accuracy of the TCS is not very good such that telescope pointing drifts
up to $\sim$1' per hour without any guiding (see Figure \ref{fig:dxdy}).
Auto guiding is thus essential both to keep the target and comparison stars
within the same field of view for a whole night and to fix the stellar positions
within several pixels to minimize systematic errors in photometry
arising from inter-pixel sensitivity variations.

Since MuSCAT2 itself has neither on-axis nor off-axis guiding camera,
we adopt a self auto-guiding system, that is, science images taken by the instrument itself
are used for guiding.
We implement a self auto-guiding software, the original version of which was developed
for MuSCAT at Okayama\cite{2015JATIS...1d5001N}\,\!.
The software calculates the stellar positional shifts between the latest science image and
a reference image in one of the four channels immediately,
and soon feeds it back to the telescope to correct the telescope position. 
Note that in the original software the stellar positional shifts are calculated
using the centroids of several bright stars, whereas this time we have introduced
a different algorithm in which one-dimensional cross correlations in both X and Y directions
are calculated between the two images,
and the shifts are calculated so as to minimize the cross correlations.
This new algorithm is more robust in the events of cloud passing and for stellar crowded fields.
The algorithm works for both on-focus and defocused observations.
The time lag between the observation of the latest image and the time that
the correction signal is sent to the telescope is $\sim$3--6 + $0.5\times T_\mathrm{exp}$ s,
where $T_\mathrm{exp}$ is the exposure time of the current image.
To avoid over correction, the guiding frequency is usually set at once in 30--60 s.
The channel to be used for guiding is user-selectable,
but usually $r$- or $i$-band channel is selected to minimize the effect of differential atmospheric refraction,
which causes a gradual positional shift of stars on a different-band channel depending on the airmass. 

Figure \ref{fig:dxdy} shows a sample of time variations of stellar position with (middle panel)
and without (top panel) the self auto-guiding software.
With guiding, long-term trends of positional shifts are significantly suppressed. 
On the other hand, stellar positions also exhibit a periodic variation in the RA direction
with the period of $\sim$169~s and the semi-amplitude of $\sim$3--6 pixels (depending on declination),
which cannot be corrected by the self auto-guiding function because of its high frequency.
Even with this periodic drift, the systematic noises in photometry arising from the inter-pixel sensitivity variations
can well be suppressed by applying Gaussian Process (see Section~\ref{subsec:precisiontest}).
However, this periodic motion also makes a stellar PSF significantly elongated when the exposure time
is longer than $\sim$5 s, which makes it difficult to observe faint objects or crowded fields with a long exposure time.

The cause of this periodic error could be attributed to inhomogeneity of gears
in the gear train that drives the RA axis of TCS. 
Tracking a sidereal object with TCS is achieved by driving a motor for the RA axis,
attached to the end of the gear train, at a constant period that corresponds to the sidereal motion.
Therefore, any inhomogeneity in one or more of the gears result in periodic acceleration
and deceleration in the telescope motion.

The ultimate solution for this problem would be to renovate the telescope control system
by attaching a high-resolution encoder directly to the RA axis and
by controlling the motor speed by referring to the encoder value.
However, it would be very costly and time-consuming.
Alternatively, as a tentative solution, we have added a new function to the telescope control system
that can sinusoidally accelerate and decelerate the motor speed so as to cancel out the tracking error. 
In this function, a new velocity of the telescope $V_\mathrm{new}$, which is the velocity that is intended
to be driven by the motor on the assumption that the gear train has no error, is given by the following formulae:

\begin{eqnarray}
V_\mathrm{new} &=& V_\mathrm{sidereal} + \delta V,\\
\delta V &=& A \sin(B x + C) + D,
\end{eqnarray}
where $V_\mathrm{sidereal}$ is the sidereal speed (15'' s$^{-1}$), $x$ is hour angle of a tracking object,
and $A$, $B$, $C$, and $D$ are coefficients.

To decide how to determine the values of coefficients, we have investigated the behavior of the periodic error
and found that (1) the period is almost constant ($\sim$169~s) over all telescope position,
(2) the period and/or the phase are slightly variable depending on hour angle,
and (3) the amplitude does not depend on hour angle but on declination.
Due to the complexity of the features (2) and (3), it is difficult to predict the values of coefficients
beforehand with sufficient precisions.
Instead, we have decided to determine or refine the coefficients during observations.
Specifically, in each observation, for the first 5 minutes the target is observed without correcting
the periodic error (but with the self guiding function activated) to gather the data of periodicity.
Subsequently, the initial values of $A$, $C$, and $D$ are determined by fitting the stellar positional data
in the RA direction gathered in the initial phase, while $B$ is fixed at a typical value.
After applying the periodic correction to the telescope tracking,
the values of $B$, $C$, and $D$ are re-evaluated every 6 minutes by using the positional data
of the last 6 minutes, where the value of $A$ is fixed at the initial value.

The bottom panel of Figure \ref{fig:dxdy} shows a sample of stellar positional data with
the periodic correction applied.
In this figure, the initial learning phase started at the time of zero, 
and the periodic correction function was applied 5 minutes later.
Note that the self auto-guiding function was also activated during this observation.
As can be seen, the dispersion of stellar position in the RA direction was gradually suppressed
after the periodic correction was applied, which finally stabilized at the level of $\sim$1.8 pixels in rms.

\subsection{Demonstration of High Precision Transit Photometry}
\label{subsec:precisiontest}

   \begin{figure}[th]
   \begin{center}
   \begin{tabular}{c}
   \includegraphics[width=16cm]{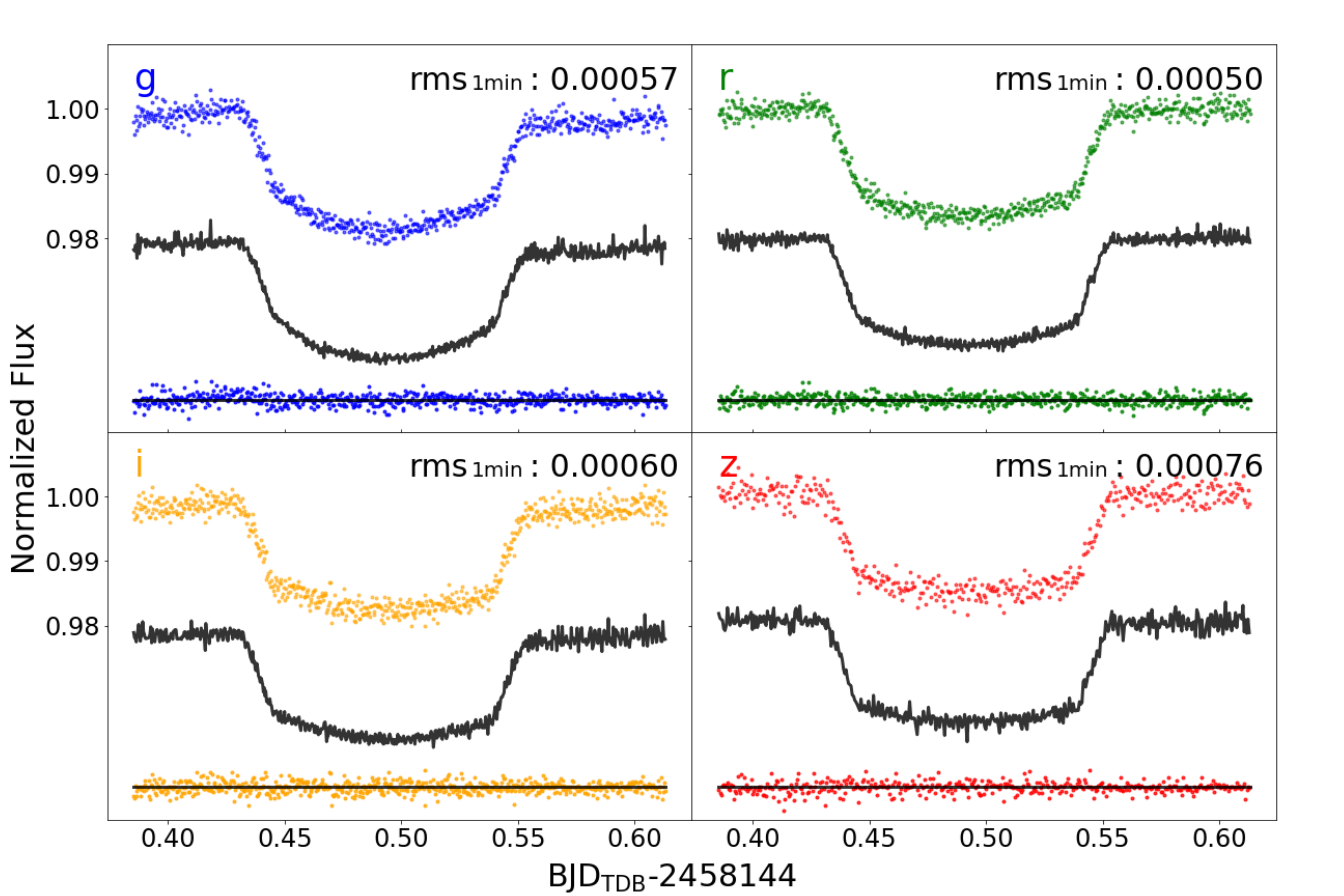}  
   \end{tabular}
   \end{center}
   \caption 
   { \label{wasp12raw} 
Transit light curves of WASP-12b taken on January 25, 2018, 
for $g$ (upper-left), $r$ (upper-right), $i$ (lower-left),
and $z_s$ (lower-right) bands, respectively.
The black solid lines indicate best-fit detrending+transit models.
The best-fit models and residuals of the data from the best-fit models are shifted for visual purposes.
Root-mean-square values of residuals with 1 min binning are shown in upper-right corner of each panel.
} 
   \end{figure} 

   \begin{figure}[th]
   \begin{center}
   \begin{tabular}{c}
   \includegraphics[width=16cm]{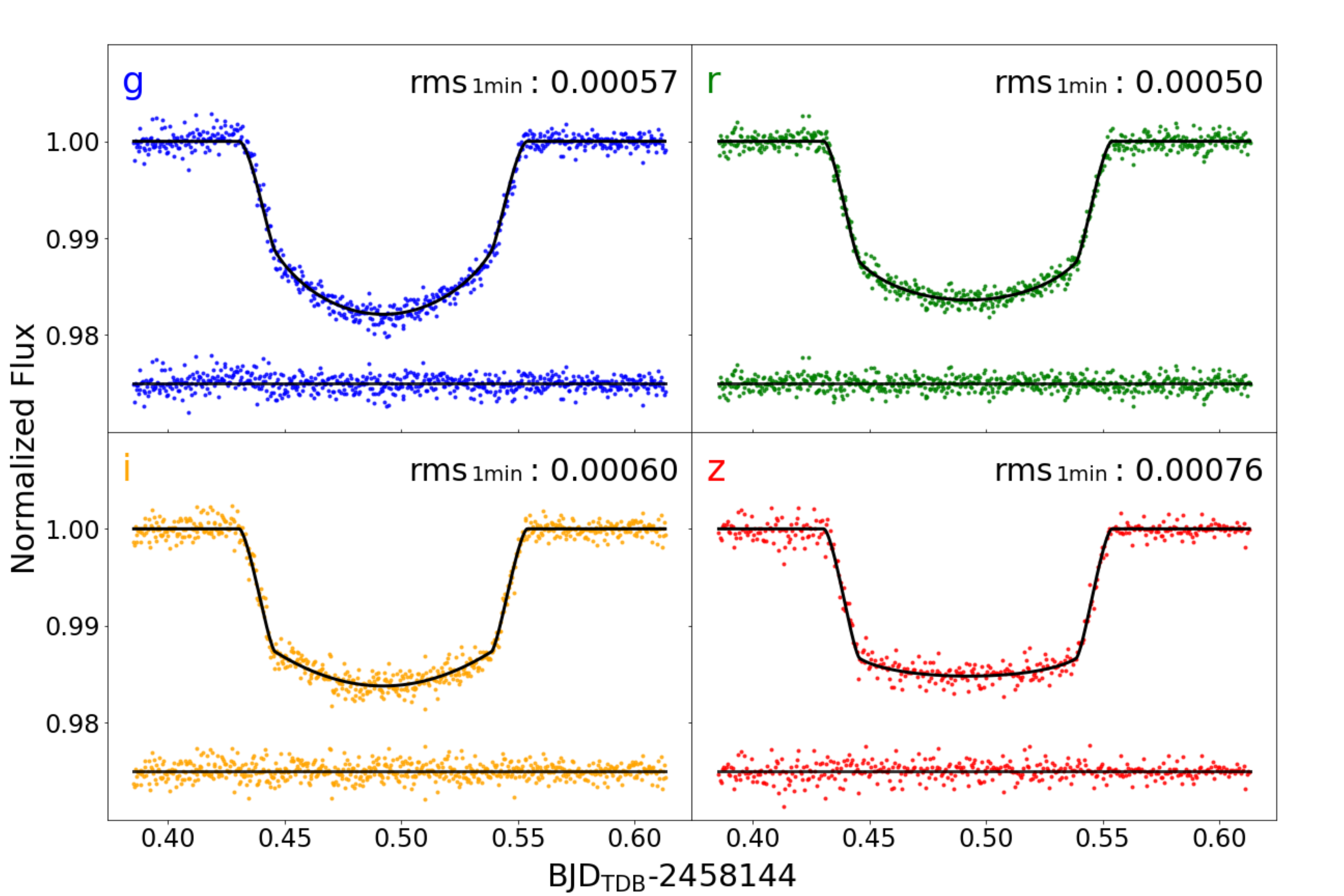}  
   \end{tabular}
   \end{center}
   \caption 
   { \label{wasp12} 
Same as Figure~\ref{wasp12raw}, but for detrended light curves.
The overlaid black solid lines indicate the best-fit transit models.
} 
   \end{figure} 

To test achievable photometric precision, we observed a full-transit of WASP-12b
on January 25, 2018.
WASP-12 is a G0 dwarf with $V=11.57\pm0.16$ mag based on the Tycho-2
catalog\cite{2000A&A...355L..27H}\,\!.
WASP-12 was observed
from 20:46 of January 25 to 02:43 of January 26 in UT
(airmass: 1.2--1.0--1.4). 
Exposure times and numbers of exposures for each band are summarized in Table \ref{tbl:testobs}.
Stellar images were defocused to avoid saturation of a brighter comparison star.
We note that because the software of correcting periodic tracking error was
under development at the time of this observation, only the self auto-guiding software
was activated to correct tracking error.

\begin{table}[t]
\begin{center}
\caption{Summary of the test observation and analysis
\label{tbl:testobs}}
\vspace{5pt}
\small
\begin{tabular}{cccccc}
\hline
\hline
Target & Band & Exp. time (s) & $N_\mathrm{data}$ $^\mathrm{a}$ & $F_\mathrm{c}/F_\mathrm{t}$ $^\mathrm{b}$ & $R_\mathrm{ap}$ $^\mathrm{c}$ (pix)   \\
\hline
WASP-12 &$g$ & 10 & 1863 & 2.8 & 16 \\
WASP-12 &$r$ & 6 & 2850 & 2.8 & 12 \\
WASP-12 &$i$ & 20 & 920 & 2.9 & 16 \\
WASP-12 &$z_{s}$ & 45 & 462 & 3.1 & 16 \\
\hline
\end{tabular}
\end{center}
{\scriptsize
$^\mathrm{a}$ Numbers of observed data points.\\
$^\mathrm{b}$ Unnormalized flux ratios of the target star and ensembles of comparison stars.\\
$^\mathrm{c}$ Applied aperture radii.\\
}
\begin{center}
\small
\caption{Rough estimates of error budgets per 1 min.
\label{tbl:error_budget}}
\vspace{5pt}
\begin{tabular}{cccccccc}
\hline
\hline
Target & Band & $\sigma_\mathrm{target}$ (\%)& $\sigma_\mathrm{comp}$ (\%)& $\sigma_\mathrm{sky}$ (\%)&
$\sigma_\mathrm{read}$ (\%)& $\sigma_\mathrm{expected}$ (\%) & rms$_\mathrm{1min}$ (\%) \\ 
\hline
WASP-12 & $g$       &  0.039 & 0.025 & 0.016 & 0.012 & 0.051 & 0.057\\
WASP-12 & $r$        &  0.034 & 0.021 & 0.010 & 0.008 & 0.043 & 0.050\\
WASP-12 & $i$        &  0.049 & 0.030 & 0.020 & 0.014 & 0.063 & 0.060\\
WASP-12 & $z_{s}$ &  0.062 & 0.037 & 0.031 & 0.016 & 0.081 & 0.076\\
\hline
\end{tabular}
\end{center}
{\footnotesize
}
\end{table}

The photometry is calculated using a Python-based photometry pipeline
specially developed for MuSCAT2.
The pipeline first computes the astrometric solution for each image frame
using an offline version of astrometry.net\cite{2010AJ....139.1782L}\,\!, and then calculates the photometry
(carrying out the basic data reductions, such as flat field division and bias subtraction)
for the target and comparison stars for a set of aperture size.
The pipeline also calculates the usual set of covariates, such as star centroids,
background sky level, airmass, etc. for each image frame to be used later in detrending.
The pipeline is optimised for defocused photometry, and calculates aperture entropy
as a proxy to the PSF FWHM, which can be used in the detrending instead of the FWHM.
The aperture entropy is calculated as,
$E = - \sum_i F_i \log F_i$,
where $F$ are the fluxes inside the photometry aperture normalized by the
total aperture flux (that is, we calculate the entropy of the flux distribution).
We also derive the effective FWHMs by calculating a table of
aperture entropies for a set of Gaussian PSFs with varying FWHM values
for all the aperture size, and use the table to map the entropies to FWHMs.

The light curves are detrended and modelled in two steps.
First, a set of optimal apertures are chosen to minimize the relative light curve point-to-point scatter,
and a transit model computed using PyTransit\cite{2015MNRAS.450.3233P}
is fitted together with a linear baseline model
with the centroid x- and y-shifts (dx and dy), sky level, airmass, and aperture entropy as covariates.
The fitting is carried out using the light curves for the four filters jointly to minimize
the sensitivity to systematics not explained by the covariates.
As a final step, we model the light curve as a sum of a transit signal and systematics represented
as a Gaussian process (GP)\cite{Rasmussen2006,2012MNRAS.419.2683G,2015ITPAM..38..252A}\,\!.
The GP uses the same covariates as the linear model,
and its hyperparameters are learned from the data by fitting a GP to the light curves
with the best-fitting transit model removed.
We use a transit model with quadratic limb darkening coefficients.
The limb darkening coefficients are free with uninformative priors so that the parameter estimates
are marginalized over all limb darkening profiles allowed by the observations.
Finally, the transit model parameter posteriors are estimated by Markov Chain Monte Carlo (MCMC)
sampling using the GP to represent the systematics.
Normalized transit light curves, best-fit detrending+transit light curve models based on the MCMC sampling,
and their residuals are plotted in Figure~\ref{wasp12raw}, and detrended transit light curves overlaid with
the best-fit transit light curve models are presented in Figure~\ref{wasp12}.

We calculate rms of residuals per 1 min for each band, and derived ones are
0.057\%, 0.050\%, 0.060\%, and 0.076\%, respectively for $g$, $r$, $i$, and $z_s$ bands.
To understand how photometric performance is achieved, 
we roughly estimate error budgets for this observation as shown in Table \ref{tbl:error_budget}.
In the table, $\sigma_\mathrm{target}$, $\sigma_\mathrm{comp}$, and $\sigma_\mathrm{sky}$
indicate typical photon noises (Poisson noises) per 1 min arising from fluxes of
the target, comparison stars, and sky-background, respectively.
$\sigma_\mathrm{read}$ is a noise caused by the readout noise (see Table~\ref{CCDsummary})
within the aperture.
Finally, $\sigma_\mathrm{expected}$ is a total expected noise per 1 min, which is calculated by
the square root of the sum of the squares of above noises.
The values of $\sigma_\mathrm{expected}$ and rms are in good agreement, meaning
the observed photometric precision can be mostly explained by photon noises of
the target, comparison stars, and sky-background noises and additional readout noise of the CCDs.
The slightly larger rms than $\sigma_\mathrm{expected}$ in $g$ and $r$ bands may be
due to a scintillation noise caused by the short exposure times in those bands.
Finally, to evaluate a possible error caused by telescope guiding, we have also tried to fit the light curves
without using dx and dy as covariates. We find that the total rms values become larger by $\sim$0.02\%
in all the bands. This means that centroid shifts may cause a potential error of $\sim$0.05-0.06\%
unless properly decorrelated.
We also note that transit parameters of the WASP-12b data will be presented
in a separated paper (Parviainen et al. in prep.).

\subsection{Discussion}
\label{subsec:discussion}

The demonstrated results can be compared with the photometric performance of MuSCAT,
which is mounted on the Okayama 1.88 m telescope.
WASP-12 was also observed with MuSCAT as a demonstration of photometric precision in
Narita~et~al.~(2015)\cite{2015JATIS...1d5001N}\,\!,
in which the photometric precision per 1 minute was 0.12\%, 0.12\%, and 0.15\%
in $g$, $r$, and $z_s$ bands, respectively.
Although one of the dominant noise sources in the MuSCAT observation was sky background,
another dominant source was unknown (unpredictable) systematics,
probably arising from telluric atmosphere (e.g., the second-order extinction\cite{2016ApJ...819...27F}).
As a result, the photometric precisions were twice worse than those achieved here with MuSCAT2,
although the expected photon noises from the target star are comparable in both cases.
As described above, the photometric noises observed with MuSCAT2 can well be explained
by known (predictable) noise sources, and no systematic noise is apparent
under favor of the higher elevation of the Teide observatory (2390 m) than the Okayama observatory (372 m).

Based on the photometric performance of MuSCAT achieved for HAT-P-14,
Fukui et al. (2016)\cite{2016ApJ...819...27F} demonstrated that MuSCAT
has an ability to probe the atmospheres of transiting planets
as small as a super-Earth/mini-Neptune ($\sim$2.5 $R_{\oplus}$) around a nearby M dwarf.
Comparing with this result, MuSCAT2 should have a similar, or even better,
ability to probe the atmosphere of exoplanets thanks to the better photometric precision
and the presence of an extra channel ($i$ band).

Multi-color transit photometry, combined with a physically-based light contamination
model and Bayesian analysis framework, can be also used to estimate the
true radius ratio of the transiting object, which allows us to distinguish
planetary-size objects from stellar-size objects.
The foundations of this approach have been laid out in previous
studies\cite{1971Icar...14...71R,2003ApJ...589.1020D,2004A&A...425.1125T}\,\!,
and the full Bayesian approach will be detailed in Parviainen et al. in prep.
MuSCAT2 will become one of the best instruments in the world for the purpose to 
discriminate whether candidates of transiting planets are real planets or false positives
due to eclipsing binaries.

We finally note how we determine a degree of defocusing and an exposure time for each target.
The optimal image size basically depends on the brightness of the target and/or comparison stars
and the dead time of the detector.
In the case of bright stars ($\lesssim$ 14 mag), sky background noise is negligible even with moderate defocusing,
while faint stars ($\gtrsim$ 15 mag) are sensitive to sky background,
which makes focused observation essential.
On the other hand, CCD cameras have finite dead time (1--4 sec),
which makes observation with a short exposure time that is comparable to the dead time inefficient.
Therefore we usually set the exposure time at 15 sec or longer for any targets except for very bright targets,
and adjust the defocused size so that the target and comparison stars will not saturate with the given exposure time.

\section{Summary}
\label{sect:summary}  

We have developed a new astronomical instrument MuSCAT2 for the TCS 1.52m telescope
in the Teide observatory, Tenerife, Spain.
MuSCAT2 has a capability of 4-color simultaneous imaging in 
$g$ (400--550 nm), $r$ (550--700 nm), $i$ (700--820 nm), and $z_{s}$ (820--920 nm) bands
with four 1k$\times$1k pixel CCDs.
The field of view of  MuSCAT2
is 7.4$\times$7.4 arcmin$^2$ with the pixel scale of 0.44 arcsec per pixel.

As shown in Section \ref{subsec:precisiontest}, MuSCAT2 is able to achieve well better precision
than 0.1\% for a G0V star with V$\sim$11.6.
The capability of demonstrated high photometric precision and 4-color simultaneous imaging
would be useful to confirm whether candidates of transiting planets
discovered by transit surveys are true planets or false positives due to eclipsing binaries.
This is especially powerful in the {\it TESS} and {\it PLATO} era, which will produce
thousands of candidates in upcoming years.
MuSCAT2 has started science operations since January 2018, and
over 250 nights per year will be allocated for the MuSCAT2 consortium at least until 2022.
MuSCAT2 will substantially contribute to follow-up transit observations of
the {\it TESS} and {\it PLATO} space missions.  

\acknowledgments 

This article is partly based on observations made with the MuSCAT2 instrument,
developed by ABC, at Telescopio Carlos S\'anchez operated on the island of Tenerife
by the IAC in the Spanish Observatorio del Teide.
This work is partly supported by JSPS KAKENHI Grant Numbers
JP18H01265, JP17H04574, JP16K13791, JP15H02063,
and JST PRESTO Grant Number JPMJPR1775.
This work is partly financed by the Spanish Ministry of Economics and Competitiveness
through grants ESP2013-48391-C4-2-R,
ESP2015-65712-C5-4-R, and AYA2015-69350-C3-2-P.


\bibliographystyle{spiejour}   


\vspace{2ex}\noindent{\bf Norio Narita} is the PI of the MuSCAT/MuSCAT2 instruments and its observing team.
He is an assistant professor of the University of Tokyo and concurrently working as a JST PRESTO researcher.

\vspace{1ex}
\noindent Biographies and photographs of the other authors are not available.


\end{spacing}
\end{document}